\documentclass[conference]{IEEEtran}
\IEEEoverridecommandlockouts

\usepackage{textcomp}
\usepackage[T1]{fontenc}
\usepackage{amsthm}
\usepackage{amsmath,amstext,amsfonts}
\usepackage{amssymb}
\usepackage{mathtools}
\usepackage[cmintegrals]{newtxmath}
\usepackage{bm}  
\usepackage{siunitx}
\usepackage{physics}
\DeclareMathOperator*{\argmin}{arg\,min}

\newtheorem{theorem}{Theorem}
\newtheorem{lemma}{Lemma}

\usepackage{graphicx}
\graphicspath{%
	{Figures/}
	{Figures/TiKz/}
	{Figures/Ipe/}
	{Figures/XFig/}
	{Figures/Inkscape/}
	{Figures/Matlab/}
	{Figures/Scanned/}
	{Figures/Saved/}
}
\usepackage{tikz}
\usepackage[list=true,labelformat=simple]{subcaption}
\usepackage{cite}
\usepackage{url}
\usepackage{xcolor}

\usepackage[inline]{enumitem}

\usepackage{algorithm,algpseudocode,float}
\usepackage{setspace} 

\algnewcommand{\algorithmicand}{\textbf{ and }}
\algnewcommand{\algorithmicor}{\textbf{ or }}
\algnewcommand{\AlgAnd}{\algorithmicand}
\algnewcommand{\AlgOr}{\algorithmicor}

\newsavebox{\myparbox}
\newlength{\myparboxwidth}

\providecommand{\norm}[1]{\lVert#1\rVert}

\newcommand{\defeq}{{\, \stackrel{\rm def}{=} \,}}

\newcommand\Angle[1]{\setbox0=\hbox{$\mskip 7mu minus 4mu#1$}%
  \raise.21ex\hbox{$/$}\hskip-0.95ex\underline{\raise\dp0\hbox{\box0}}}

\ExplSyntaxOn

\NewDocumentCommand \vect { s o m }
 {
  \IfBooleanTF {#1}
   { \vectaux*{#3} }
   { \IfValueTF {#2} { \vectaux[#2]{#3} } { \vectaux{#3} } }
 }

\DeclarePairedDelimiterX \vectaux [1] {\lbrack} {\rbrack}
 { \, \dbacc_vect:n { #1 } \, }

\cs_new_protected:Npn \dbacc_vect:n #1
 {
  \seq_set_split:Nnn \l_tmpa_seq { , } { #1 }
  \seq_use:Nn \l_tmpa_seq { \enspace }
 }
\ExplSyntaxOff

\usepackage[firstpageonly=true]{draftwatermark}
\usepackage{hyperref}
\hypersetup{
    colorlinks=true,
    linkcolor=blue,
    filecolor=magenta,
    urlcolor=cyan,
    bookmarks=true,
}
\usepackage[noabbrev]{cleveref}  
\Crefname{figure}{Fig.}{Figs.}

\def\BibTeX{{\rm B\kern-.05em{\sc i\kern-.025em b}\kern-.08em
    T\kern-.1667em\lower.7ex\hbox{E}\kern-.125emX}}
\begin{document}

\title{A Data-Driven Model-Reference Adaptive Control Approach Based on Reinforcement Learning
\thanks{This work was partially supported by NSERC Grant~EGP~537568-2018.}
}

\author{%
  \IEEEauthorblockN{Mohammed Abouheaf\IEEEauthorrefmark{1}, Wail Gueaieb\IEEEauthorrefmark{2}, Davide Spinello\IEEEauthorrefmark{3} and Salah Al-Sharhan\IEEEauthorrefmark{4}}
  \IEEEauthorblockA{\IEEEauthorrefmark{1}College of Technology, Architecture \& Applied Engineering\\
    Bowling Green State University, Bowling Green, 43402, OH, USA\\
    Email: mabouhe@bgsu.edu}
  \IEEEauthorblockA{\IEEEauthorrefmark{2}School of Electrical Engineering and Computer Science\\
    \IEEEauthorrefmark{3}Department of Mechanical Engineering\\
    University of Ottawa, Ottawa, Ontario, Canada K1N~6N5\\
    Email: \{wgueaieb, dspinell\}@uottawa.ca}
  \IEEEauthorblockA{\IEEEauthorrefmark{4}Machine Intelligence Research Labs\\
    Auburn, Washington 98071-2259, USA\\
    Email: salah27@ieee.org}
}

\maketitle

\begin{abstract}
Model-reference adaptive systems refer to a consortium of techniques that guide plants to track desired reference trajectories. Approaches based on theories like Lyapunov, sliding surfaces, and backstepping are typically employed to advise adaptive control strategies.  
The resulting solutions are often challenged by the complexity of the reference model and those of the derived control strategies. 
Additionally, the explicit dependence of the control strategies on the process dynamics and reference dynamical models may contribute in degrading their efficiency in the face of uncertain or unknown dynamics.
A model-reference adaptive solution is developed here for autonomous systems where it solves the Hamilton-Jacobi-Bellman equation of an error-based structure. 
The proposed approach describes the process with an integral temporal difference equation and solves it using an integral reinforcement learning mechanism.     
This is done in real-time without knowing or employing the dynamics of either the process or reference model in the control strategies. 
A class of aircraft is adopted to validate the proposed technique. 
\end{abstract}

\begin{IEEEkeywords}
Model-Reference Control, Integral Bellman Equation, Integral Reinforcement Learning, Adaptive Critics
\end{IEEEkeywords}

\DraftwatermarkOptions{%
angle=0,
hpos=0.5\paperwidth,
vpos=0.97\paperheight,
fontsize=0.012\paperwidth,
color={[gray]{0.2}},
text={
  \newcommand{\thispaperdoi}{10.1109/ROSE52750.2021.9611772}
  \newcommand{\thispaperCopyrightYear}{2021}
  \parbox{0.99\textwidth}{This is the postscript version of the published paper. (doi: \href{http://dx.doi.org/\thispaperdoi}{\thispaperdoi})\\
    \copyright~\thispaperCopyrightYear~IEEE.  Personal use of this material is permitted.  Permission from IEEE must be obtained for all other uses, in any current or future media, including reprinting/republishing this material for advertising or promotional purposes, creating new collective works, for resale or redistribution to servers or lists, or reuse of any copyrighted component of this work in other works.}},
}

\section{Introduction}
Optimal tracking control problems for linear time-invariant systems are typically solved using the Linear Quadratic Tracking Regulator~(LQT)~\cite{Lewis2012}. This is done offline via solving a set of differential equations backward-in-time. The guidance and tracking problems cover a wide range of autonomous and intelligent applications~\cite{aastrom2013adaptive}. Some of which require sophisticated modeling to describe the dynamics, that when available allow to develop model-reference adaptive mechanisms that are robust and effective in real-world applications. Model-reference adaptive control employs theories in the class of sliding surfaces, backstepping, and Lyapunov-based methods to develop online adaptive strategies~\cite{Vempaty2016,hu2010distributed,Chen2021}. These methods, being based on systems' models, inherit their inherent complexity, uncertainties and approximations. Additionally, in networked multi-agent systems, the order of the underlying individual dynamics along with the network's connectivity may further raise the complexity of the problem.

Herein, a model-reference adaptive control~(MRAC) approach, also known as model reference adaptive system~(MRAS), is designed to guide dynamic systems using an online integral reinforcement learning (IRL) mechanism, where the stability is guaranteed for a broad range of slowly varying dynamics under standard conditions that apply to a large class of systems.
%
%
The sums of squared error-based optimization approach is employed to design an adaptive control system in~\cite{Moore2014} for underactuated systems of moderate dimensions (less than 12).
An MRAS is developed to stablize the lateral motion of a 5-DOF car-trailer
system in~\cite{Vempaty2016}, where the yaw rates of the car and trailer are assigned by using a Lyapunov stability criterion.
An adaptive switch controller is developed to control a hydraulic actuator system in~\cite{Zuo2017}. It coordinates among a PID control unit and MRAS to improve the tracking performance and hence the response time.
In~\cite{SHI2017}, a robust MRAS based on augmented error dynamics is developed where solving a set of  linear matrix inequalities results in the error feedback matrix. This approach tackled the high oscillations in the actuator response due to the tuning with high gain adaptive rates.
A decentralized MRAS is developed in~\cite{BenAmor2017} for large-scale interconnected systems, such as robotic manipulators.
An adaptive control approach combined with anti-windup structure is developed for autonomous underwater vehicles (AUVs) in~\cite{Sarhadi2016}.
Another augmented MRAS with anti-windup commentator is applied to tackle the limitations caused by the actuator saturation for a class of aircraft in~\cite{Chen2018}.
An adaptive control scheme is developed for uncertain hypersonic flight vehicles in~\cite{Liu2018}. A barrier function technique is employed to advise a novel handling of the angle-of-attack of this aircraft. 
An MRAC approach is employed to design an adaptive cruise system for vehicles in~\cite{Adamu2020}.
A combined approach that uses $L_1$ adaptive control and a backstepping method is employed to achieve fast adaptation and dynamic positioning of an underactuated surface vessel in~\cite{Chen2021}. It uses a nonlinear observer to estimate the uncertainties and to guarantee a chattering-free performance.  
A leader-follower tracking problem is solved using distributed estimators under uncertain dynamical environment in~\cite{hu2010distributed}. The work used the sliding surface approach to coordinate among the different tasks.
An adaptive mechanism is employed to trace the peaks of unknown fields for a mulit-agent system in~\cite{jadaliha2010adaptive}.
Pinning control ideas are employed in~\cite{AbouheafAuto14,AbouheafCDC13,AbouheafIJCNN13} to leader-followers tracking problems with multi-agent systems under a fixed graph topology.

The approximation features of machine learning (ML) can be applied to solve adaptive control problems~\cite{sut92,Sutton_1998}. Such approaches have been adopted in several applications ranging from the autonomous control of aircraft and underactuated sea vessels to regulating distributed generation sources~\cite{AbouheafTrans20,AbouheafSMC20,Abouheafmag2021,AbouheafIRL2019}.  
Reinforcement learning (RL) provides the agent with a mechanism to learn the best control strategy-to-follow~\cite{Sutton_1998}. This is done according to multiple interactions between the agent and the environment to explore how to transit to a better state. The attempted policies are either penalized or rewarded in order to maximize the sum of a cumulative reward~\cite{Sutton_1998,Bertsekas1996}.   
The RL approaches are solved using two-step techniques like policy iteration (PI) and value iteration (VI)~\cite{Bertsekas1996,Busoniu2010,Abouheapolicy2017}.
The Integral Reinforcement Learning approaches have been used to solve control problems with nonlinear-performance indices~\cite{Abouheaf2019}.
An adaptive critics is usually built as a neural network platform to approximate the strategy-to-follow using an actor neural network and the associated value which is modeled by a critic neural network~\cite{Bertsekas1996}. These two-step approaches are employed to solve a class of cooperative control problems for multi-agent systems communicating over graphs~\cite{AbouheafCTT2015,AbouheafAuto14}.  
An adaptive Fuzzy-RL is adopted to control the flocking motion of a swarm of robots in~\cite{ICRA21}.
The PI solutions are implemented using regression models like least squares and batch least squares~\cite{Busoniu2010, Srivastava2019}.

The main contribution of this work is the development of a generalized data-driven MRAC solution that is independent of the plant dynamics. 
It approximates an optimal control Hamiltonian structure to find an integral temporal difference expression. Then, it solves the underlying integral Bellman equation using IRL in real-time.
The proposed solution is flexible in terms of the system order, unlike those associated with solving the optimal tracking problems of high-order systems~\cite{Lewis2012,Bahare14}. The algorithm runs online without explicitly solving the underlying optimal tracking problem.        
The remaining parts of the paper are arranged as follows: \Cref{Prob} formulates the model-reference adaptive problem. The duality between the optimal control and adaptive control ideas is explained in \Cref{Opt}. \Cref{IRL} introduces the IRL implementation of the adaptive control solution using a Value Iteration approach. \Cref{Anl} shows the simulation results for an autonomous aircraft system.  Finally, some concluding remarks are presented in \Cref{Conc}.

\section{Problem Formulation}
\label{Prob}
The MRAS decides on the control strategy-to-follow along the trajectory of the reference model dynamics~\cite{aastrom2013adaptive}. This is done based on minimizing the error between the output $\bm{y}(t)\in \mathbb{R}^P$ of the process and that of the reference model $\bm{y}^{ref}(t)\in \mathbb{R}^P$. Herein, it is assumed that the dynamics of the process is linearizable into a state-space model in the form 
\begin{align}
  \label{eq:Sys}
  \dot{\bm{X}} & = \bm{A} \, \bm{X}+ \bm{B} \, \bm{u} 
  & \bm{y} & = \bm{C} \, \bm{X}
\end{align}
where ${\bm{X}}\in\mathbb{R}^n$ and  ${\bm{u}}\in\mathbb{R}^m$ are the states and control input signals of the aircraft, respectively. The drift dynamics, control input, and output measurement matrices are denoted by the unknown matrices ${\bm{A}}\in\mathbb{R}^{n \times n},$   ${\bm{B}}\in\mathbb{R}^{n\times m},$ and ${\bm{C}}\in\mathbb{R}^{p\times n}$, respectively.

\begin{figure}[htb]
  \centering
  \includegraphics[width=1\linewidth]{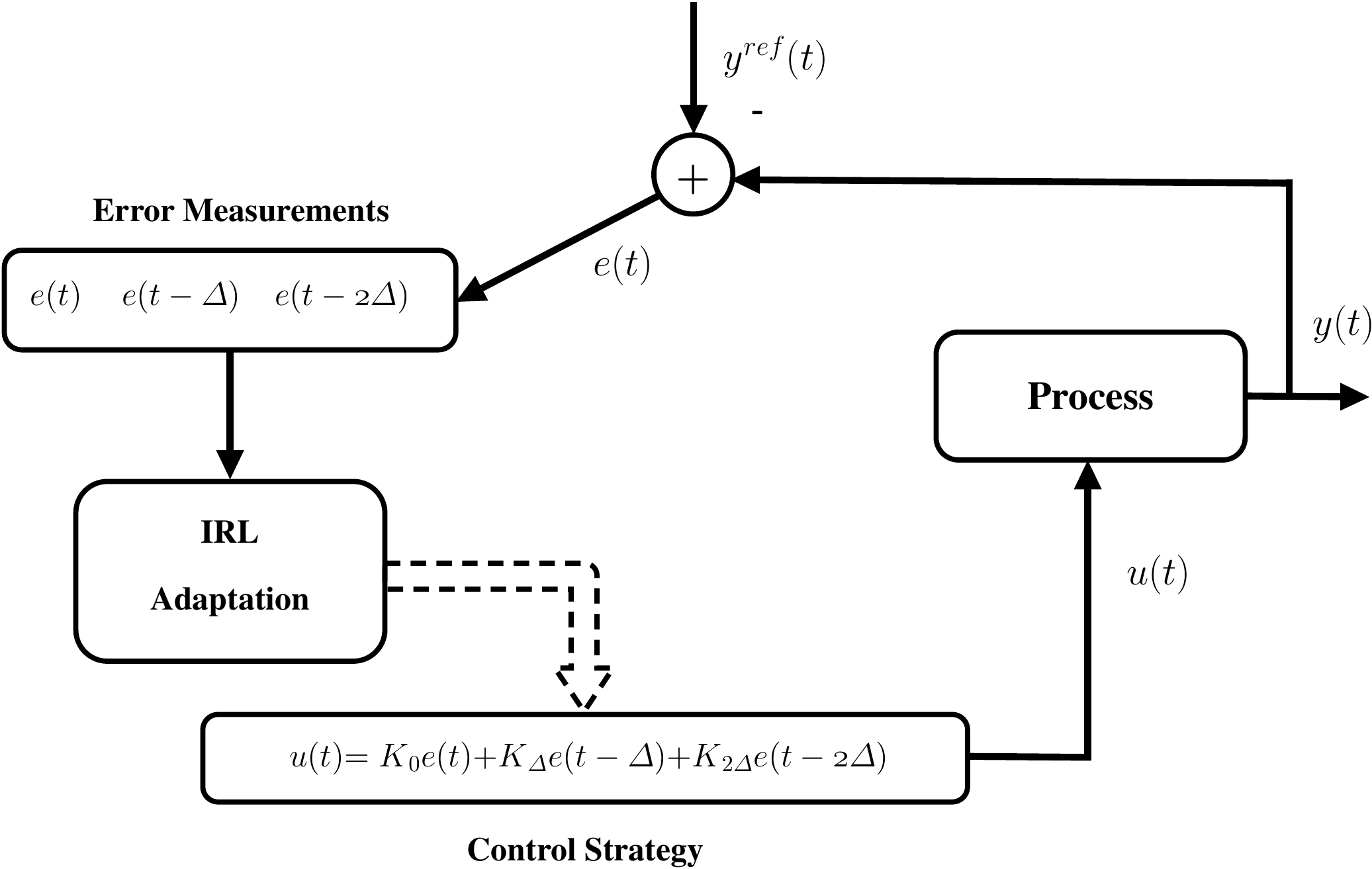}
  \caption{Model-reference adaptive system}
  \label{fig:control}
\end{figure}

The objective of the optimization problem and hence the adaptive controller is to decide on the adaptive strategy $\pi$ defined by three control gain matrices $\{K_0,K_\Delta,K_{2\Delta}\}$ of proper dimensions, with $\bm{u}^{\pi}=K_0\,e(t)+K_\Delta\,e(t-\Delta)+K_{2\Delta}\,e(t-2\Delta)$ along the reference trajectory in order to  minimize the error $\bm{e}(t)=\bm{y}(t)-\bm{y}^{ref}(t)$; i.e.,~$\lim_{t \to \infty} \norm{ \bm{e}(t) } = 0$.
%
%
The symbol $\Delta$ refers to a sampling time interval. The adaptive control solution works in a different way when compared to the optimal control one. It is concerned with finding a control law to minimize the tracking errors in real-time without targeting the optimal control gains. In this work, a single output system is considered (i.e., $p=1$). The solution provided herein combines an optimal control framework with an adaptive control mechanism that is based on IRL to solve the MRAC problem as shown in \Cref{fig:control}. It is worth to note that, the proposed adaptive strategy depends on only the reference signal $y^{ref}(t)$ and the process output $y(t)$. It does not rely on the state-space matrices.

\section{Optimal and Adaptive Control}
\label{Opt}
This section discusses the optimal control framework and hence the underlying adaptive system. The structure of the control strategy depends on the history of error samples, i.e.,~$t, \, t-\Delta, \, t-2\Delta$, etc. The width of the time window to consider is a design parameter which is left to the discretion of the designer, which gives flexibility to the proposed approach. A wider time window may lead to a higher performance but at the expense of computational complexity. Herein, a time window of 3 samples is chosen, i.e.,~$t, \, t-\Delta, \, t-2\Delta$, which yields a second-order error dynamic system. This is unlike many classical adaptive structures which are mostly based on zero- or first-order error strategies~\cite{aastrom2013adaptive,Lewis2012,Chen2018}.  
The error data are stacked together in a vector 
$\bm{E}(t) = \vect{ e(t) , e(t-\Delta) , e(t-2\Delta) }^T \in \mathbb{R}^3$.
A performance index $J^{\pi}(t)$ is used to measure the quality of the applied strategy $u^\pi(t)$ by evaluating a cost or utility function $U$ over the infinite horizon, such that
\begin{equation*}
  J^{\pi}(t)=\int_{t}^{\infty} U(\bm{E}(\tau),\bm{u}^\pi(\tau)) \, d \tau,
\end{equation*}
where
$U(\bm{E}(\tau),\bm{u}^\pi(\tau))=\frac{1}{2}\left(\bm{E}^T(\tau)\, \bm{Q}\, \bm{E}(\tau)+{\bm{u}^{\pi}}^T(\tau)\, \bm{R}\, \bm{u}^\pi(\tau)\right)$. The weighting matrices are denoted by $\bm{0} < \bm{Q}\in\mathbb{R}^{3\times3}$ (positive definite) and $0 < \bm{R}\in\mathbb{R}$.

The mathematical development of the adaptive control solution combines together ideas from optimal control, adaptive control, and IRL, as is detailed next.

\begin{theorem}
  \label{thm:CTHJB}
  Let $V \left( \bm{E}(t),\bm{u}^\pi(t) \right)$ be a nonnegative multi-input single-output quadratic convex function which is nil only at the origin. Then,  
  \begin{enumerate}[leftmargin=*, widest*=9]
  \item $V\left(\bm{E}(t),\bm{u}^\pi(t)\right)$ satisfies a continuous-time Hamilton-Jacobi-Bellman (CTHJB) equation $H(\bm{E}(t),\grad{V}\left(\bm{E}(t),\bm{u}^{\pi*}(t)\right), \bm{u}^{\pi*}(t))=0$.
  \item $V\left(\bm{E}(t),\bm{u}^\pi(t)\right)$ is a Lyapunov function.
  \end{enumerate}	
\end{theorem}

\begin{IEEEproof}
\begin{enumerate}[leftmargin=*, widest*=9]
\item The continuous-time Hamiltonian function for the optimal control problem is given by
\begin{equation*}
H(\bm{E}(t),\bm{\lambda}(t), \bm{u}^\pi(t))=\bm{\lambda}^T(t) \, \dot{\bm{Z}}^\pi(t)+U\left(\bm{E}(t),\bm{u}^\pi(t)\right),
\end{equation*}
where $\bm{\lambda}$ is a Lagrange multiplier associated with the constraint $\dot{\bm{Z}}^\pi(t)=0$ with 
$\bm{Z}^\pi(t)=\vect{ \bm{E}^T(t) , {\bm{u}^\pi(t)}^T }^T$.\\
Let the value function $V$ be quadratic and convex in $\bm{Z}^\pi(t)$, such that
\begin{equation}
  V\left(\bm{E}(t),\bm{u}^\pi(t)\right)
  =\frac{1}{2}
  {\bm{Z}^\pi}^T(t)\,
  \bm{H}\,
  \bm{Z}^\pi(t),
  \label{val}
\end{equation}
where
$\displaystyle
\bm{H}
\equiv
\begin{bmatrix*}[l]
  \bm{H}_{\bm{E}\bm{E}}
  & 
  \bm{H}_{\bm{E}\bm{u}}
  \\
  \bm{H}_{\bm{u}\bm{E}}
  &
  \bm{H}_{\bm{u}\bm{u}}
\end{bmatrix*}
\in \mathbb{R}^{4\times4} > \bm{0},
$
$\boldsymbol{H}_{\bm{u}\bm{u}} \in \mathbb{R}$, and $\boldsymbol{H}_{\bm{u}\bm{E}} \in \mathbb{R}^{1 \times 3}$.
The Hamilton-Jacobi theory provides a relation between Lagrange multiplier $\bm{\lambda}$ and the value function $V\left(\bm{E}(t),\bm{u}^\pi(t)\right)$~\cite{Lewis2012},
where $\bm{\lambda} =\grad{V} \left(\bm{E}(t),\bm{u}^\pi(t)\right)={\partial V\left(\bm{E}(t),\bm{u}^\pi(t)\right)}/{\partial \bm{Z}^\pi(t)}$. Therefore, this value function satisfies the continuous-time Bellman equation given by
\begin{equation}
  \grad{V} \left(\bm{E}(t),\bm{u}^\pi(t)\right)^T \, \dot{\bm{Z}}^\pi(t)+U\left(\bm{E}(t),\bm{u}^\pi(t)\right)=0. 
  \label{CTBell}
\end{equation}
This expression is an infinitesimal equivalent of $J^{\pi}(t)\defeq V\left(\bm{E}(t),\bm{u}^\pi(t)\right)=\int_{t}^{\infty} U(\bm{E}(\tau),\bm{u}^\pi(\tau)) \, d \tau$. Hence, \eqref{CTBell} is equivalent to
\begin{multline}
  H(\bm{E}(t),\grad{V}\left(\bm{E}(t),\bm{u}^{\pi}(t)\right), \bm{u}^{\pi}(t))= \\
  \dot V\left(\bm{E}(t),\bm{u}^\pi(t)\right)+U\left(\bm{E}(t),\bm{u}^\pi(t)\right)=0. 
  \label{CTBelldt}
\end{multline}
Applying the Bellman stationarity condition yields the optimal strategy ${u^{\pi*}(t)}$ which is calculated as ${u^{\pi*}(t)}=\argmin_{u^\pi(t)}H(\bm{E}(t),\grad{V}\left(\bm{E}(t),\bm{u}^{\pi}(t)\right), \bm{u}^{\pi}(t))$). This yields the CTHJB 
\begin{multline}
  H(\bm{E}(t),\grad{V}\left(\bm{E}(t),\bm{u}^{\pi*}(t)\right), \bm{u}^{\pi*}(t))= \\
  \dot V^{*}\left(\bm{E}(t),\bm{u}^{\pi*}(t)\right)+U\left(\bm{E}(t),\bm{u}^{\pi*}(t)\right)=0, 
  \label{CTHJB}
\end{multline}
where $V^{*}$ is the optimal solution of the CTHJB equation.
\item The value function $V$ is a Lyapunov candidate function. Its derivative according to \eqref{CTBelldt} is
$\dot V\left(\bm{E}(t),\bm{u}^\pi(t)\right)=-U\left(\bm{E}(t),\bm{u}^\pi(t)\right) \le 0$.
Thus, $V\left(\bm{E}(t),\bm{u}^\pi(t)\right)$ is a Lyapunov function.
\end{enumerate}
\end{IEEEproof}

The following result explains how to apply \Cref{thm:CTHJB} to come up with a temporal difference form (the integral Bellman equation) in order to adapt the control gains $\{K_0,K_\Delta,K_{2\Delta}\}$ in real-time.

\begin{theorem}
  \label{thm:IRL-Bellman}
  The value function $V^*\left(\bm{E}(t),\bm{u}^\pi(t)\right)$ satisfies the following IRL-Bellman equation
  \begin{multline}
    \label{IRLOpt}
    V^*\left(\bm{E}(t),\bm{u}^{\pi*}(t)\right)=\int_{t}^{t+\Delta}U\left(\bm{E}(\tau),\bm{u}^{\pi*}(\tau)\right)\,d\tau
    \\
    +
    V^*\left(\bm{E}(t+\Delta),\bm{u}^{\pi*}(t+\Delta)\right)
  \end{multline}
  where  the optimal control gains $\{K^*_0,K^*_\Delta,K^*_{2\Delta}\}$ are decided using Bellman optimality principles. 
\end{theorem}

\begin{IEEEproof}
  Bellman equation~\eqref{CTBell} may be discretized using Euler approach, such that
\begin{equation*}
  \frac{V\left(\bm{E}(t),\bm{u}^{\pi}(t)\right)-V\left(\bm{E}(t+\Delta),\bm{u}^{\pi}(t+\Delta)\right)}{\Delta}
  =U\left(\bm{E}(t),\bm{u}^{\pi}(t)\right)
\end{equation*}
which can be rearranged to
\begin{multline}
  V\left(\bm{E}(t),\bm{u}^{\pi}(t)\right)=\int_{t}^{t+\Delta}U\left(\bm{E}(\tau),\bm{u}^{\pi}(\tau)\right)\,d\tau
  \\
  +
  V\left(\bm{E}(t+\Delta),\bm{u}^{\pi}(t+\Delta)\right).
  \label{IRLBell}
\end{multline}
This equation is referred to as the Integral Bellman equation, which is equivalent to~\eqref{CTBell}.  Applying Bellman optimality principle yields the optimal strategy  $\bm{u}^{\pi*}(t)=\argmin_{\bm{u}^{\pi}(t)} V\left(\bm{E}(t),\bm{u}^{\pi}(t)\right)$. Using~\eqref{val} yields
\begin{equation}
\bm{u}^{\pi*}(t)=- \boldsymbol{H}_{\bm{u}\bm{u}}^{-1}  \boldsymbol{H}_{\bm{u}\bm{E}} \, \bm{E}(t).
\label{optpol}
\end{equation}
It is noted that this control strategy is model-free because it does not rely on the process dynamics. Hence, the optimal strategy ${\pi*}$ is defined by ${\pi*}=- \boldsymbol{H}_{\bm{u}\bm{u}}^{-1}  \boldsymbol{H}_{\bm{u}\bm{E}}$  and the optimal control gains $\bm{K}^*= \vect{ K^*_0 , K^*_\Delta , K^*_{2\Delta} } = - \boldsymbol{H}_{\bm{u}\bm{u}}^{-1} \, \boldsymbol{H}_{\bm{u}\bm{E}}$.
Applying this optimal policy into the integral Bellman equation~\eqref{IRLBell} yields the Integral Bellman optimality form~\eqref{IRLOpt}.
\end{IEEEproof}

The next Lemma lays down the stability conditions of the suggested control law.

\begin{lemma}
  \label{thm:stability}
  Let $ V\left(\bm{E}(0),\bm{u}^\pi(0)\right)\le\cal{M}$ for an upper bound $\cal{M}$. Then, $\lim_{t\rightarrow \infty} \dot V\left(\bm{E}(t),\bm{u}^\pi(t)\right)=0$ and the error system is asymptotically stable (i.e.,~$\lim_{t\rightarrow  \infty} e(t) = 0$) for a bounded reference signal  $\bm{y}^{ref}(t)$.
\end{lemma} 

\begin{IEEEproof}
$V\left(\bm{E}(t),\bm{u}^\pi(t)\right)$ is a Lyapunov function. Therefore, $ V\left(\bm{E}(t),\bm{u}^\pi(t)\right)\le V\left(\bm{E}(0),\bm{u}^\pi(0)\right)\le\cal{M}$. This implies that, $V\left(\bm{E}(t),\bm{u}^\pi(t)\right)\in L_\infty$. So, $e(t),e(t-\Delta),e(t-2\Delta) \in L_\infty$ and $\bm{H}$ (or equivalently $\bm{K}^*$) $\in L_{\infty}$. 
Taking into consideration that $\int_{0}^{t}\dot V\left(\bm{E}(\vartheta),\bm{u}^\pi(\vartheta)\right) \, d\vartheta=V\left(\bm{E}(t),\bm{u}^\pi(t)\right)-V\left(\bm{E}(0),\bm{u}^\pi(0)\right)$ and that $V$ is a Lyapunov function, we get $-\int_{0}^{t}\dot V\left(\bm{E}(\vartheta),\bm{u}^\pi(\vartheta)\right) \, d\vartheta \le V\left(\bm{E}(0),\bm{u}^\pi(0)\right)$. This yields $\dot V\left(\bm{E}(t),\bm{u}^\pi(t)\right) \in L_\infty$, where the error dynamics $\dot e(t), \, \dot e(t-\Delta), \, \dot e(t-2\Delta) \in L_\infty$. Using the CTHJB equation~\eqref{CTHJB} and the strategy~\eqref{optpol} we infer that
$\int_{0}^{t} \bm{E}^T(\vartheta) \, \left(\bm{Q}+\bm{K}^{*T}\bm{R}\bm{K}^*\right)\,\bm{E}(\vartheta) \, d\vartheta \le V\left(\bm{E}(0),\bm{u}^\pi(0)\right)$, which reveals that $e(t), \, e(t-\Delta), \, e(t-2\Delta) \in L_2$ and $\dot V\left(\bm{E}(t),\bm{u}^\pi(t)\right) \in L_2$. Following Barbalat's lemma~\cite{aastrom2013adaptive}, $\lim_{t\rightarrow \infty} \dot V\left(\bm{E}(t),\bm{u}^\pi(t)\right)=0$ and $\lim_{t\rightarrow \infty} e(t)=0$, which proves the asymptotic stability of the system.
\end{IEEEproof}
This shows that the choice of the value function~\eqref{val} and the model-free strategy~\eqref{optpol} lead to an optimal solution for the Integral Bellman optimality equation~\eqref{IRLOpt}.

\section{Integral Reinforcement Learning Solution}
\label{IRL}
The control gains of the MRAC are selected using an online IRL approach that employs the error vector $\bm{E}(t)$ measured in real-time. This is done by solving IRL-Bellman optimality equation~\eqref{IRLOpt} with the model-free optimal policy~\eqref{optpol} simultaneously using a Value-Iteration method in real-time. The adaptive control solution requires a real-time adaptation mechanism to solve this system of equations. Therefore, means of adaptive critics are used to implement the online solution of the adaptive control problem. The next Theorem highlights the convergence results of the IRL-Value Iteration process.

\begin{theorem}
  \label{thm:convergence}
  Consider the value function~\eqref{val} and the control policy~\eqref{optpol}. Then,
  \begin{enumerate}[leftmargin=*, widest*=9]
  \item The value function update law~\eqref{IRLOpt} yields a non-decreasing positive definite sequence $0 \le V^0 \le V^1 \le V^2 \le \dots \le V^*$, which asymptotically converges to the solution of the IRL Bellman optimality equation~\eqref{IRLOpt}.
  \item The policies generated by~\eqref{optpol} are stabilizing policies.
  \end{enumerate}
\end{theorem}
        
\begin{IEEEproof}
\begin{enumerate}[leftmargin=*, widest*=9]
\item According to \Cref{thm:stability}, the initial value function is bounded so that $0 < V^0\left(\bm{E}(\bm{0}),\bm{u}^\pi(\bm{0})\right)\le\cal{M}$ and the IRL Bellman expression~\eqref{IRLBell} employing a strategy $\pi$ yields $V^{r+1}\left(\bm{E}(t),\bm{u}^{\pi}(t)\right)=\int_{t}^{t+\Delta}U\left(\bm{E}(\tau),\bm{u}^{\pi}(\tau)\right)\,d\tau
+V^{r}\left(\bm{E}(t+\Delta),\bm{u}^{\pi}(t+\Delta)\right)$, where $r$ is a value iteration update index. 
Hence, $\forall r$,
$
 V^{r+1}\left(\bm{E}(t),\bm{u}^{\pi}(t)\right)=\sum_{i=0}^{r}V^{1}\left(\bm{E}(t+i\,\Delta),\bm{u}^{\pi}(t+i\, \Delta)\right)
-\sum_{i=1}^{r}V^{0}\left(\bm{E}(t+i\,\Delta),\bm{u}^{\pi}(t+i\, \Delta)\right)
$. This leads to
$V^{r+1}\left(\bm{E}(t),\bm{u}^{\pi}(t)\right)=\int_{t}^{t+r\,\Delta} U\left(\bm{E}(\tau),\bm{u}^{\pi}(\tau)\right) d\tau
+V^{0}\left(\bm{E}(t+(r+1)\,\Delta),\bm{u}^{\pi}(t+(r+1)\, \Delta)\right)$. 
Therefore, the value sequence is non-decreasing, i.e.,~ $0 \le V^0\le V^1\le \dots \le V^r\le V^{r+1}$, $\forall r$.
According to \Cref{thm:stability}, the error system is asymptotically stable and hence the cost is bounded so that $0<\int_{0}^{\infty} U\left(\bm{E}(\tau),\bm{u}^{\pi}(\tau)\right) d\tau \le \cal{U}$. The former sequence of value functions is upper bounded by $\cal{M+U}$. Then, the value function sequence $V^r$ satisfies $ 0 \le V^0\le V^1\le V^2\le\dots \le V^*$ and asymptotically converges to the optimal solution $V^*$ of~\eqref{IRLOpt}.
\item The optimal policies $\bm{u}^r$, $\forall r, \Delta$, satisfy the following IRL-Bellman optimality equation $V^{r}\left(\bm{E}(t),\bm{u}^{r}(t)\right)-V^{r}\left(\bm{E}(t+\,\Delta),\bm{u}^{r}(t+\, \Delta)\right)=\int_{t}^{t+\Delta} U\left(\bm{E}(\tau),\bm{u}^{r}(\tau)\right) d\tau
$. Using the stability result of \Cref{thm:stability} and employing the value function $V^r(\dots)$, or equivalently the policy $\bm{u}^r$, over the infinite horizon, we get
$V^{r}\left(\bm{E}(t),\bm{u}^{r}(t)\right) \ge V^{r}\left(\bm{E}(t+\,\Delta),\bm{u}^{r}(t+\, \Delta)\right)\ge V^{r}\left(\bm{E}(t+2\,\Delta),\bm{u}^{r}(t+2\, \Delta)\right)\ge\dots\ge \lim_{\tau\rightarrow \infty}V^{r}\left(\bm{E}(\tau),\bm{u}^{r}(\tau)\right)=0$. Thus, the strategies $u^r$, $\forall r$, are stabilizing and hence admissible. Therefore, $V^{r}\left(\bm{E}(t+r\Delta),\bm{u}^{r}(t+r\Delta)\right) \ge \dots \ge V^{2}\left(\bm{E}(t+2\Delta),\bm{u}^{2}(t+2\Delta)\right)\ge V^{1}\left(\bm{E}(t+\,\Delta),\bm{u}^{1}(t+\, \Delta)\right)\ge V^{0}\left(\bm{E}(t),\bm{u}^{0}(t)\right)$.
\end{enumerate}
\end{IEEEproof}

The adaptive critics approximates the value of some strategy $\pi$  using a critic network and the associated optimal strategy using an actor network. The critic network is motivated by the structure of the value function $V$, so that
\begin{equation*}
\hat V\left(\bm{E}(t),\hat{\bm{u}}^\pi(t)\right)
=\frac{1}{2}
{\bm{I}^\pi}^T(t)\,
\bm{\Theta}_c\,
\bm{I}^\pi(t),
\end{equation*}
where $\bm{I}^\pi(t)= \vect{ \bm{E}(t) , \hat{\bm{u}}^\pi(t) }^T$ and ${\hat{\bm{u}}^\pi(t)}$ is the control signal using the approximated strategy. The adaptive critic adaptation weights are set as $\bm{\Theta}_c = \bm{H} >0$. 

Similarly, the actor network approximation is given by
\begin{equation*}
\hat{\bm{u}}^\pi(t)=\bm{\Theta}_a \, \bm{E}(t),
\end{equation*}
where $\bm{\Theta}_a = \vect{ K_0 , K_\Delta , K_{2\Delta} }$ represents the actor adaptation weights.  

In order to adapt the MRAC weights (i.e.,~the actor and critic weights), a gradient descent approach is considered. Thus, the critic and actor adaptation errors are calculated with the help of~\eqref{IRLOpt} and~\eqref{optpol} which represent the optimal control solution.
The critic adaptation error is given by
$\varepsilon^{Critic}= \frac{1}{2}\left(\hat V\left(\bm{E}(t),\hat{\bm{u}}^\pi(t)\right)-\tilde V(t)\right)^2$
where 
$\tilde V(t)=U\left(\bm{E}(t),\hat{\bm{u}}^\pi(t)\right)+\hat V\left(\bm{E}(t),\hat{\bm{u}}^\pi(t)\right)$. Thus, the critic adaptation law follows
\begin{equation}
\bm{\Theta}^{(r+1)}_c=\bm{\Theta}^{(r)}_c-\zeta_c \, \varepsilon^{Critic(r)} \bm{I}^\pi \,{\bm{I}^\pi}^T,
\label{Crt_Law}
\end{equation}
where $0<\zeta_c<1$ is a critic-adaptation rate.

In a similar fashion, the actor adaptation rule follows a gradient descent approach where the actor adaptation error is given by  
$\varepsilon^{Actor}= \frac{1}{2}\left(\hat{\bm{u}}^\pi-\tilde u\right)^2$
with $\tilde u= - \boldsymbol{\Theta}_{c\hat{\bm{u}}\hat{\bm{u}}}^{-1} \, \boldsymbol{\Theta}_{c\hat{\bm{u}}\bm{E}}$ which is motivated by the solution structure~\eqref{optpol}. Hence, the resulting update law for the actor weights is given by
 \begin{equation}
 \bm{\Theta}^{(r+1)}_a=\bm{\Theta}^{(r)}_a-\zeta_a \, \varepsilon^{Actor(r)}  \,{\bm{E}}^T
 \label{Act_Law}
\end{equation}
where $0<\zeta_a<1$ is an actor-adaptation rate.
The synthesis of the adaptive control solution is detailed in \Cref{alg:alg1}.

\begin{algorithm}[htb!]
  \setstretch{1} 
  \caption{Model-Reference Adaptive Control Algorithm}\label{alg:alg1}
  \begin{algorithmic}[1] 
    \Require
    \Statex Number of adaption steps ${\cal N}$
    \Statex Adaptation step ${\Delta}$
    \Statex Actor and critic adaptation rates $\zeta_a$ and $\zeta_c$
    \Statex Weighting matrices $\boldsymbol{Q}$ and $\bm{R}$
    \Statex Convergence threshold $\delta$ and a time window width $\cal L$
    \Ensure
    \Statex Adapted gains $\bm{\Theta}^{(t+k\Delta)}_a$ and $\bm{\Theta}^{(t+k\Delta)}_c,\quad k=0,1,\dots, \cal{N}$
    \Statex
    \State Initialize $\bm{E}(0)$, $\bm{\Theta}^{(0)}_a$ and $\bm{\Theta}^{(0)}_c$\Comment{to zero, for example}
    \State Get the initial reference signal $\bm{y}^{ref}(0)$
    \State $k \gets 0$ 
    \State Weights\_converged $\gets$ false
    \While {$k <{\cal N}$ \AlgAnd Weights\_converged $=$ false}
    \State Calculate the control signal $\hat{\bm{u}}(k\Delta)$, then apply it to the process to get $\bm{y}((k+1)\, \Delta)$ 
    \State Use previous step and $\bm{y}^{ref}((k+1)\, \Delta)$ to calculate $e(t+(k+1)\,\Delta)$
    \State Find $V(\bm{E}(k\,\Delta),\hat{\bm{u}}(k\Delta))$ and $U(\bm{E}(k\,\Delta),\hat{\bm{u}}(k\Delta))$
    \State Get $\bm{E}((k+1)\,\Delta)$ and $\hat{\bm{u}}((k+1)\,\Delta)$, and deduce $V(\bm{E}((k+1)\,\Delta,\hat{\bm{u}}((k+1)\Delta))$
    \State Compute $\tilde V(k\,\Delta)$ and $\tilde{\bm{u}}(k\Delta)$
    \State Update the critic weights $\Theta_c^{(k+1)}$\Comment{using~\eqref{Crt_Law}}
    \State Update the actor weights $\Theta_a^{(k+1)}$\Comment{using~\eqref{Act_Law}}
    \If{$k > {\cal L}$\AlgAnd $\norm{\bm{\Theta}_c^{ (k+1-l)}-\bm{\Theta}_c^{(k-l)}} \le \delta,$ $\forall l \in \{0,1,\ldots,{\cal L}\}$,}
    \State $\bm{\Theta}_c^{(*)} \gets \bm{\Theta}_c^{(k+1)}$ 
    \State Weights\_converged $\gets$ true
    \EndIf
    \State $k \leftarrow k+1$
    \EndWhile
    \Statex \Return $\bm{\Theta}^{(t+k\Delta)}_a$ and $\bm{\Theta}^{(t+k\Delta)}_c$, for $k=0,1,\dots, \cal{N}$
  \end{algorithmic}
\end{algorithm}

It is worth to note that, this actor-critic adaptation mechanism provides an adaptive control solution framework that is based on an optimal control architecture. Further, it gives a flexibility to select the actor form to approximate the optimal strategy. It is different from the classical control structures in the way Lyapunov value functions are used. Herein, it is used to find an integral temporal difference structure that can be solved using IRL. This results in a straightforward equation that is easy to manipulate and solve. On the other hand, unlike similar algorithms in the literature, the actor-critic adaption structures, represented by~\eqref{Crt_Law} and~\eqref{Act_Law}, do not depend on the process model. As a result, the proposed development provide easy-to-implement features of the IRL-based adaptive critics mechanism.

\section{Results}
\label{Anl}
The adaptive model-free controller in \Cref{alg:alg1} is applied for the control of the longitudinal motion of an aircraft. The underlying longitudinal states of the aircraft (making the vector $\bm{X}$ in~\eqref{eq:Sys}) are the aircraft's angle of attack $\alpha$ and the pitch rate $q$. The control signal $\bm{u}$ is the elevator deflection, while the output $\bm{y}$ is the vertical acceleration. The control scheme does not require the a priori knowledge of the process dynamics. Nevertheless, to simulate the aircraft behavior, the following approximate linearized model is adopted~\cite{Chen2018}:
\begin{align*}
  \bm{A}
  & =
    \begin{bmatrix*}[l]
      -8.76 & 0.954
      \\
      -177 & -9.92
    \end{bmatrix*}
      &
      \bm{B}
      & =
        \begin{bmatrix*}[l]
          -0.697
          \\
          -168
        \end{bmatrix*}
  &
    \bm{C}
      & =
        \begin{bmatrix*}[l]
          -0.8 & -0.04
        \end{bmatrix*}
\end{align*}
The simulation and learning parameters of the MRAS are taken as $\Delta=\SI{0.1}{\s}$, $\zeta_c=0.5$, and $\zeta_a=0.5$. The weighting matrices of the cost function are selected, such that $\bm{Q}=I_{3 \times 3}$ (identity matrix) and $\bm{R}=1$. The remaining parameters are set to ${\cal{N}}=180$, ${\cal{L}}=10$, and $\delta=10^{-8}$.

\begin{figure*}[h!]
  \centering
  \subcaptionbox{Reference/measured vertical acceleration%
    \label{fig:ref_1}}
  {%
    \includegraphics[width=0.32\textwidth]{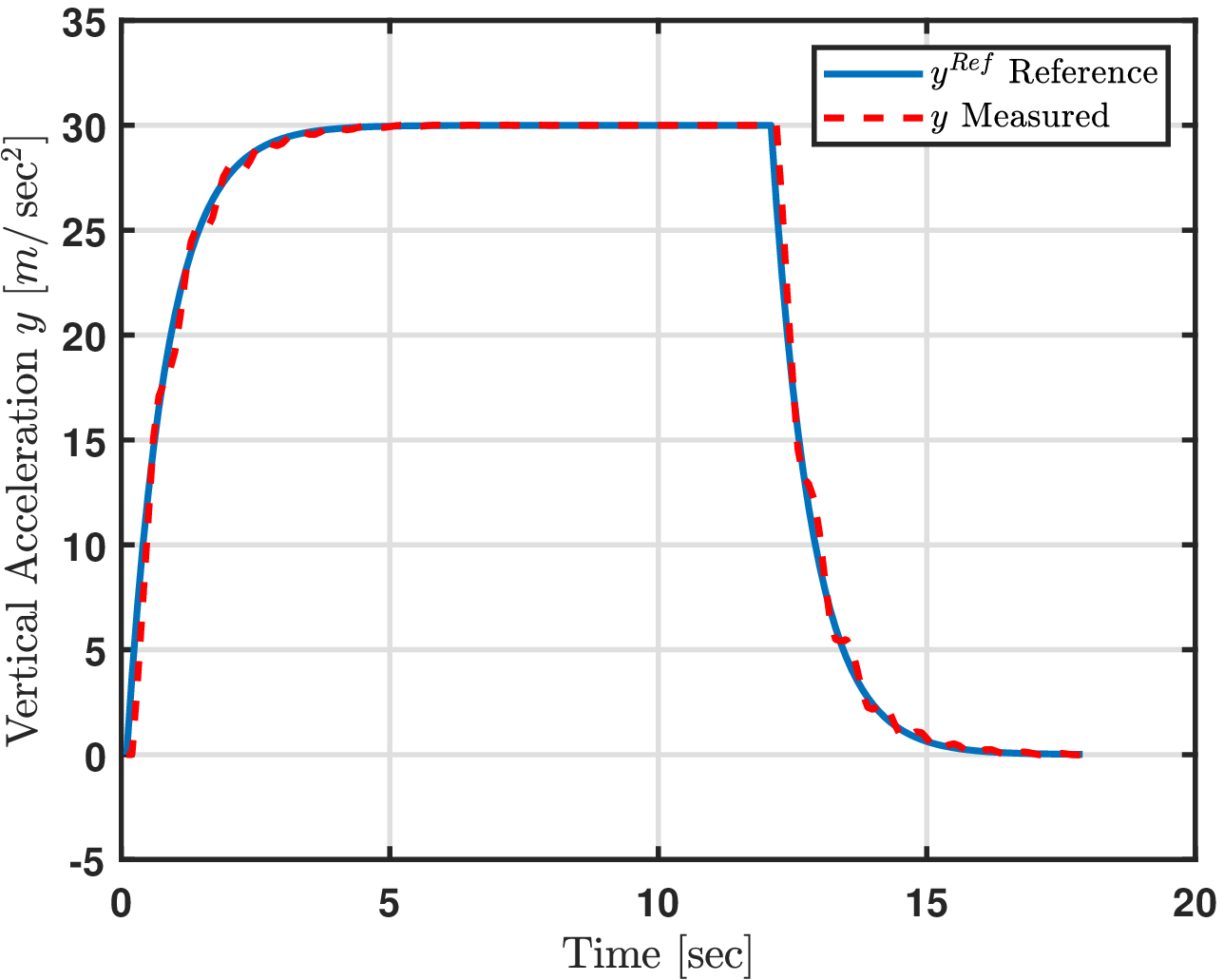}%
  }
  \hfill
  \subcaptionbox{Tracking error~$e(t)$%
    \label{fig:err_1}}
  {%
    \includegraphics[width=0.32\textwidth]{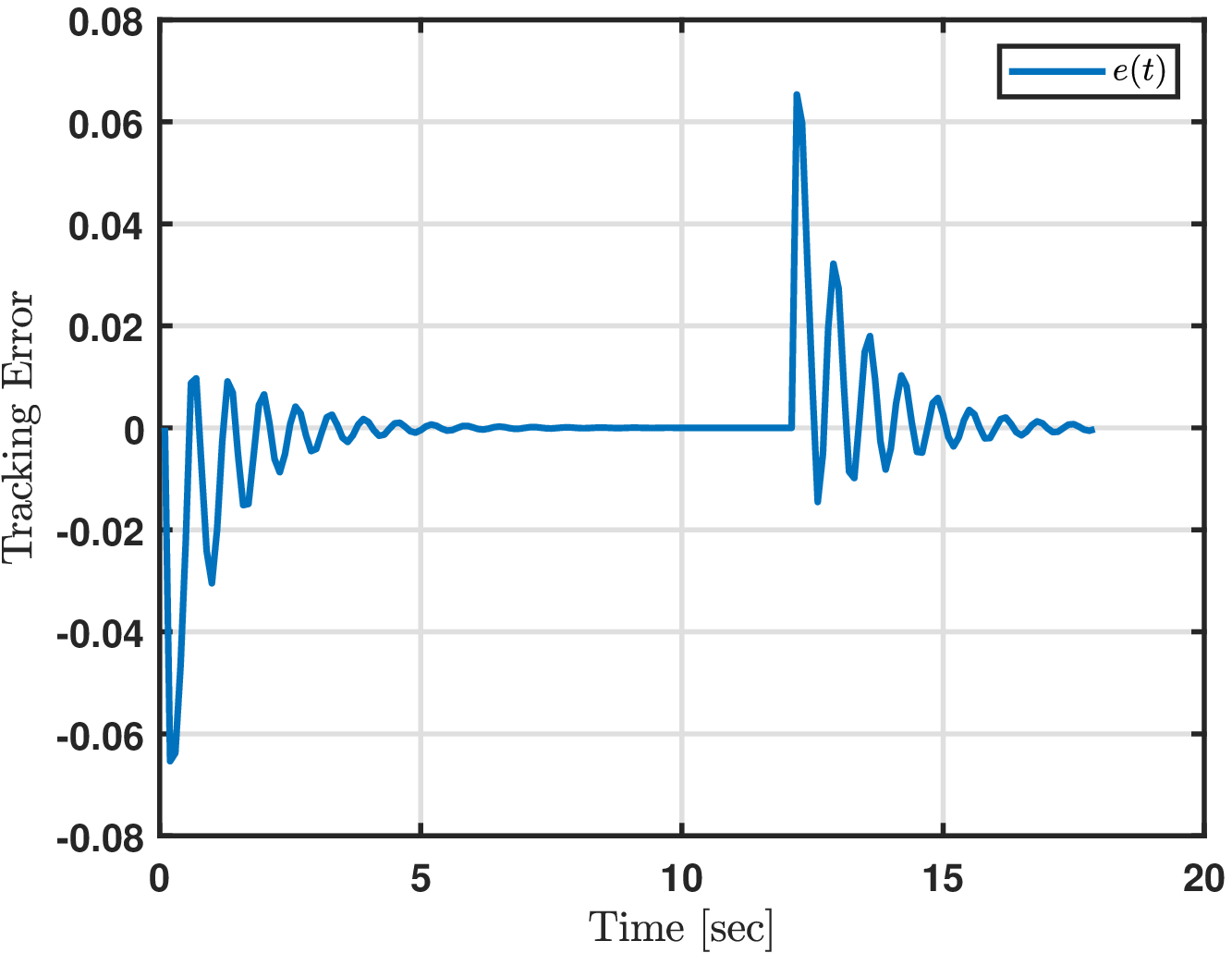}%
  }
  \hfill
  \subcaptionbox{System states $\alpha(t)$ and~$q(t)$%
    \label{fig:stat_1}}
  {%
    \includegraphics[width=0.32\textwidth]{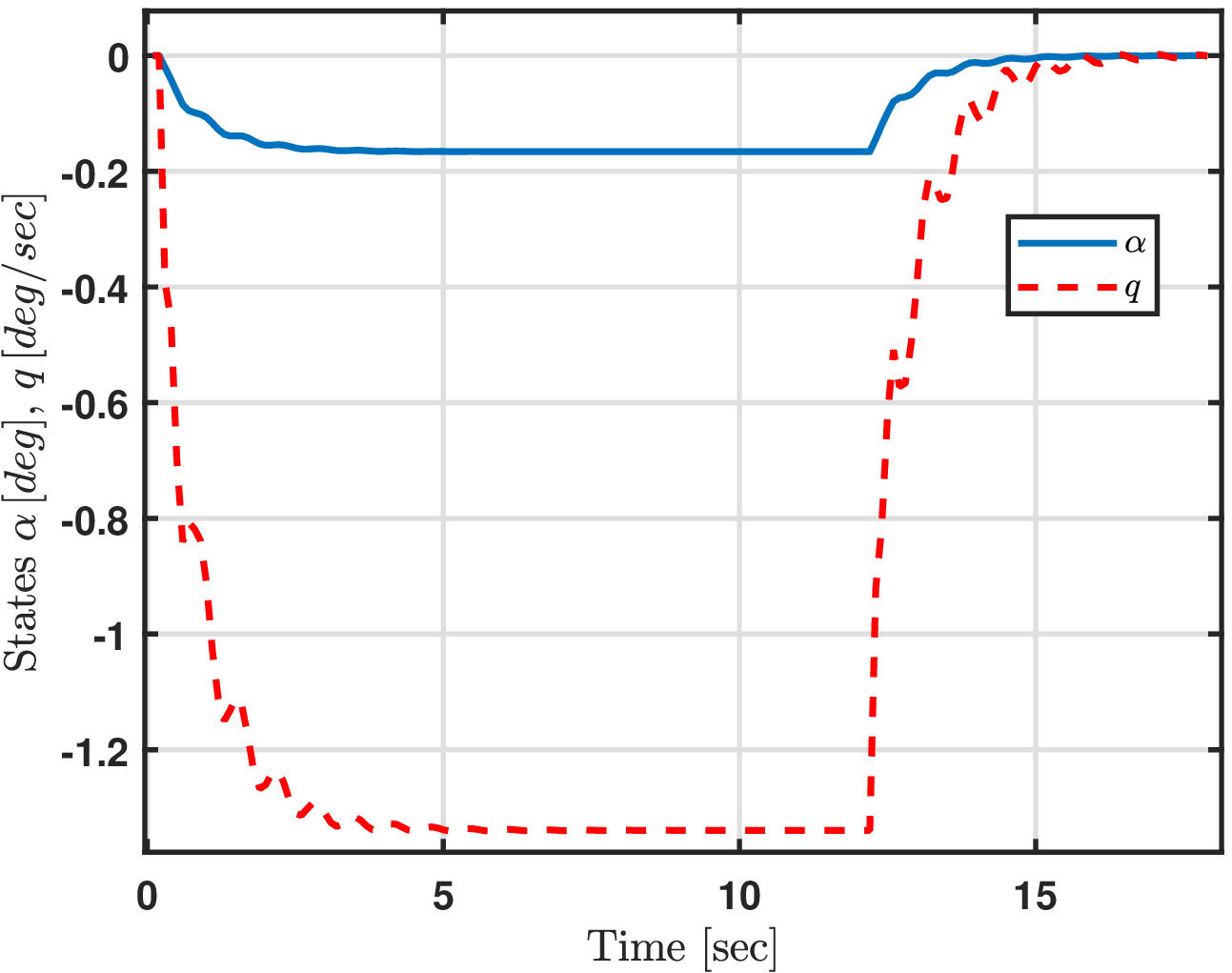}%
  }
  \\[2ex]
  \mbox{}\hfill
  \subcaptionbox{Elevator deflection~$\bm{u}(t)$%
    \label{fig:con_1}}
  {%
    \includegraphics[width=0.32\textwidth]{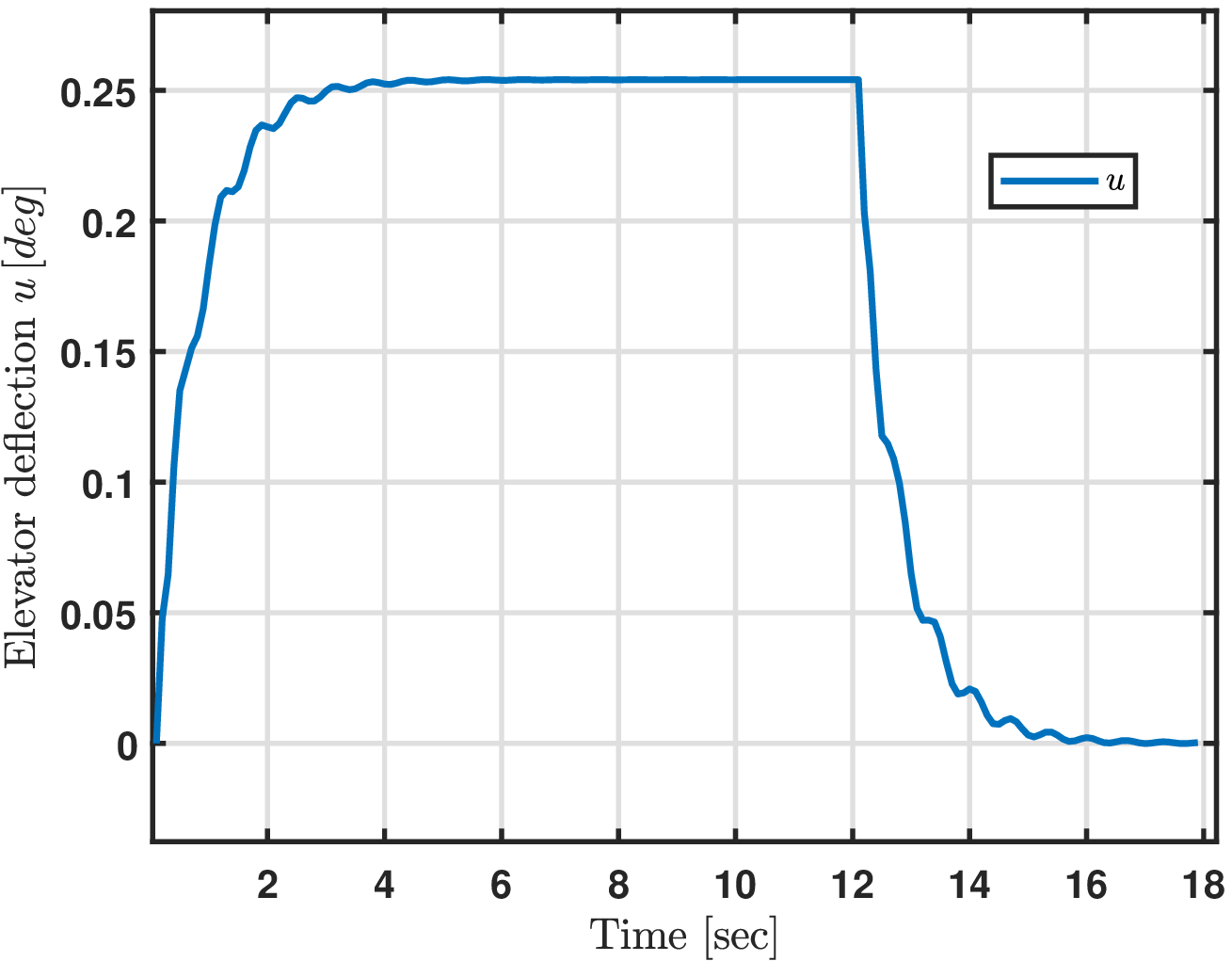}%
  }
  \hfill
  \subcaptionbox{Control gains~$K(t)$%
    \label{fig:act_1}}
  {%
    \includegraphics[width=0.32\textwidth]{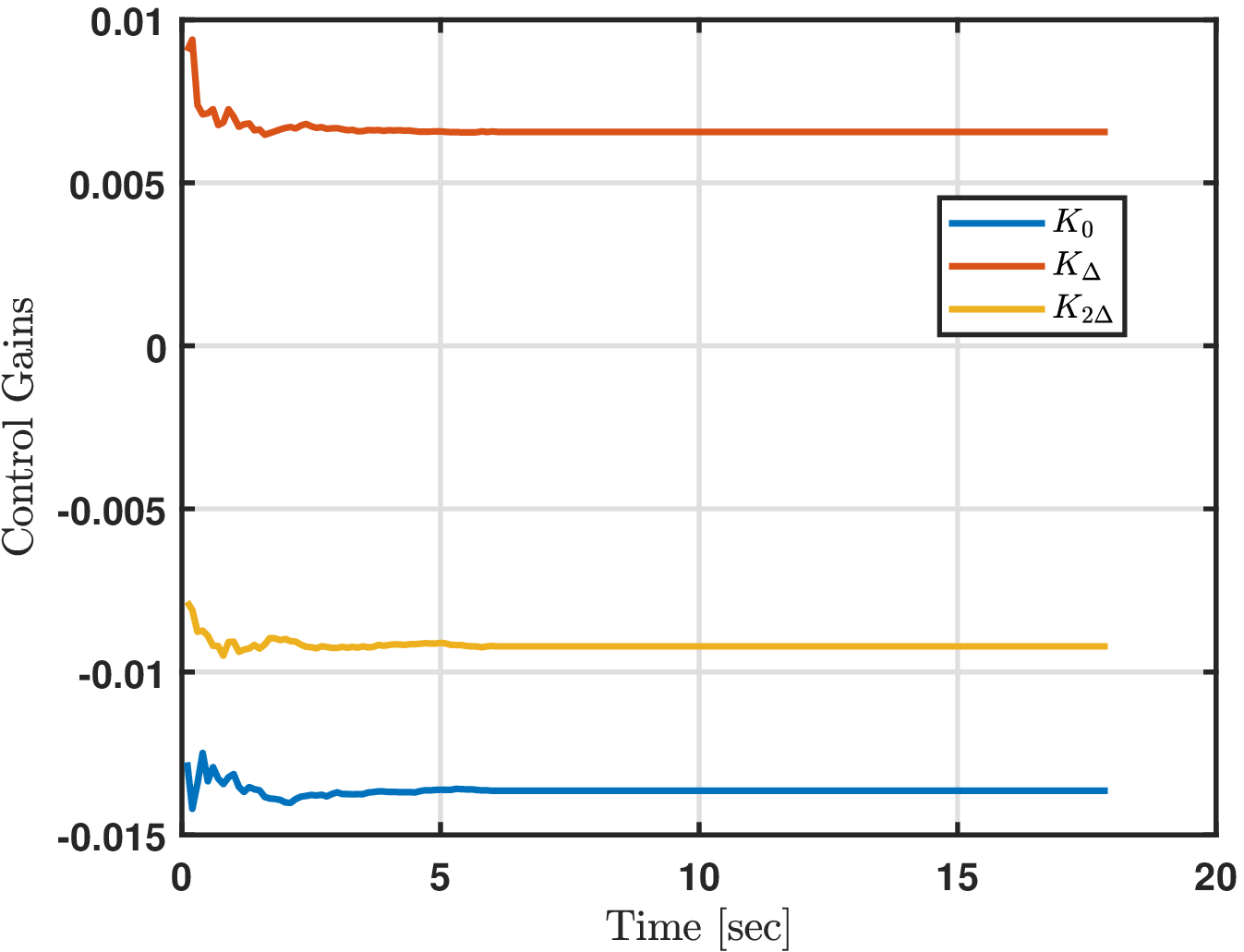}%
  }
  \hfill\mbox{}
  \caption{Simulation results of test case~1\label{fig:main 1}} 
\end{figure*}

Two simulation cases are considered. The first employs a reference-model of a filtered square wave signal with an amplitude of $\SI{30}{\meter/\s^2}$. The outcome of this first test case is revealed in \Cref{fig:main 1}. As expected, the tracking error is asymptotically annihilated and the control gains are driven to converge (see \Cref{fig:err_1,fig:act_1}). 

\begin{figure*}[h!]
  \centering
  \subcaptionbox{Reference/measured vertical acceleration%
    \label{fig:ref_2}}
  {%
    \includegraphics[width=0.32\textwidth]{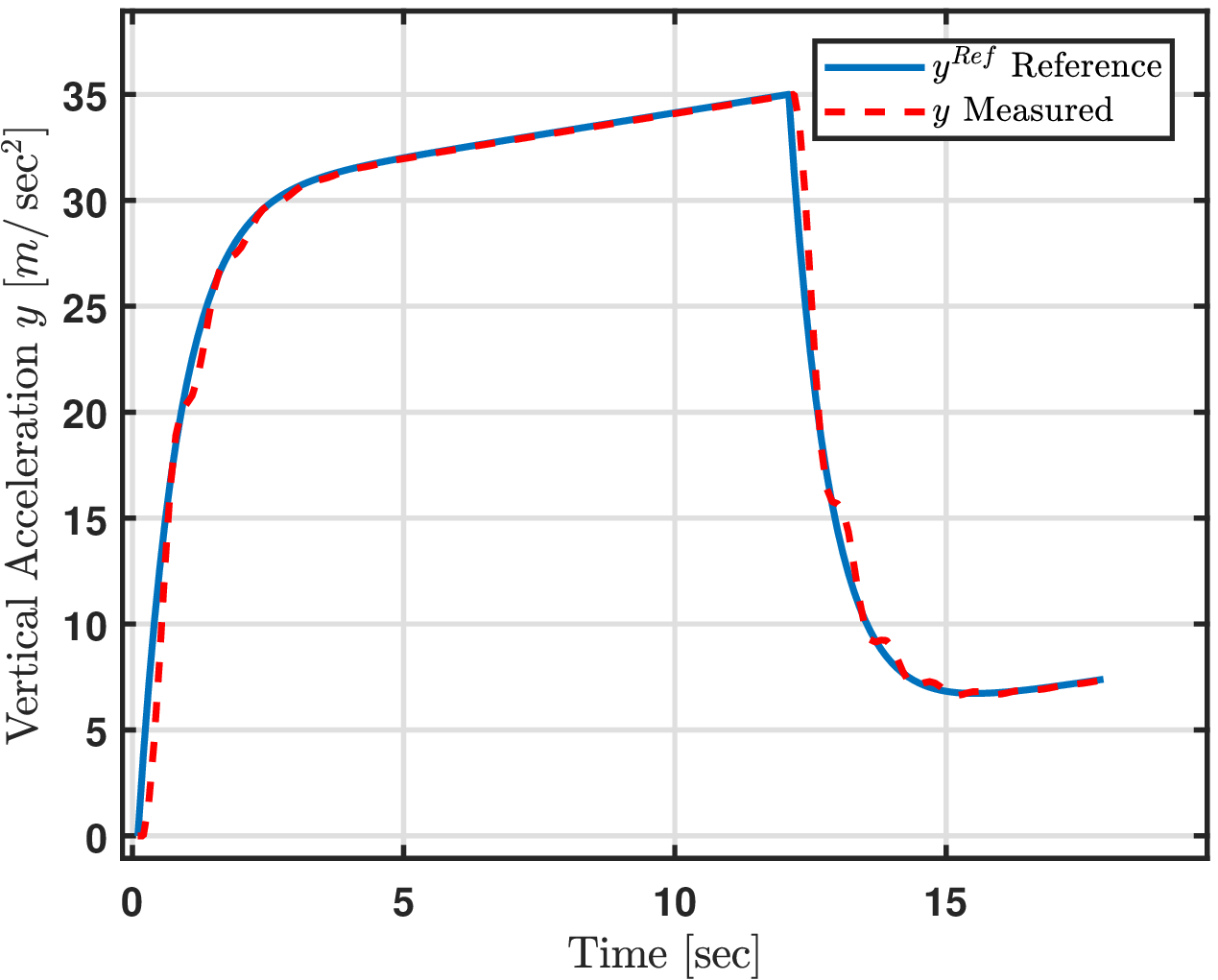}%
  }
  \hfill
  \subcaptionbox{Tracking error~$e(t)$%
    \label{fig:err_2}}
  {%
    \includegraphics[width=0.32\textwidth]{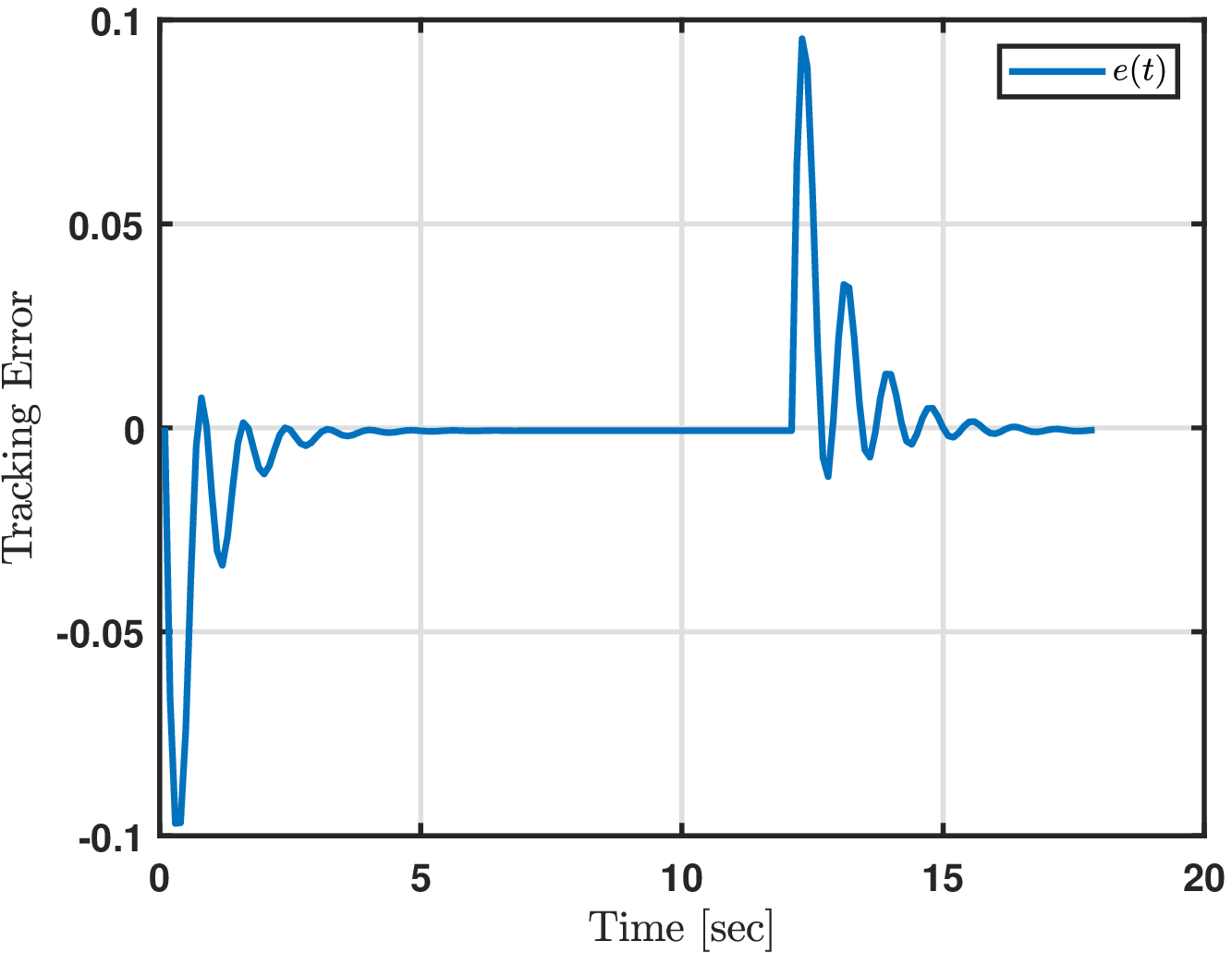}%
  }
  \hfill
  \subcaptionbox{System states $\alpha(t)$ and~$q(t)$%
    \label{fig:stat_2}}
  {%
    \includegraphics[width=0.32\textwidth]{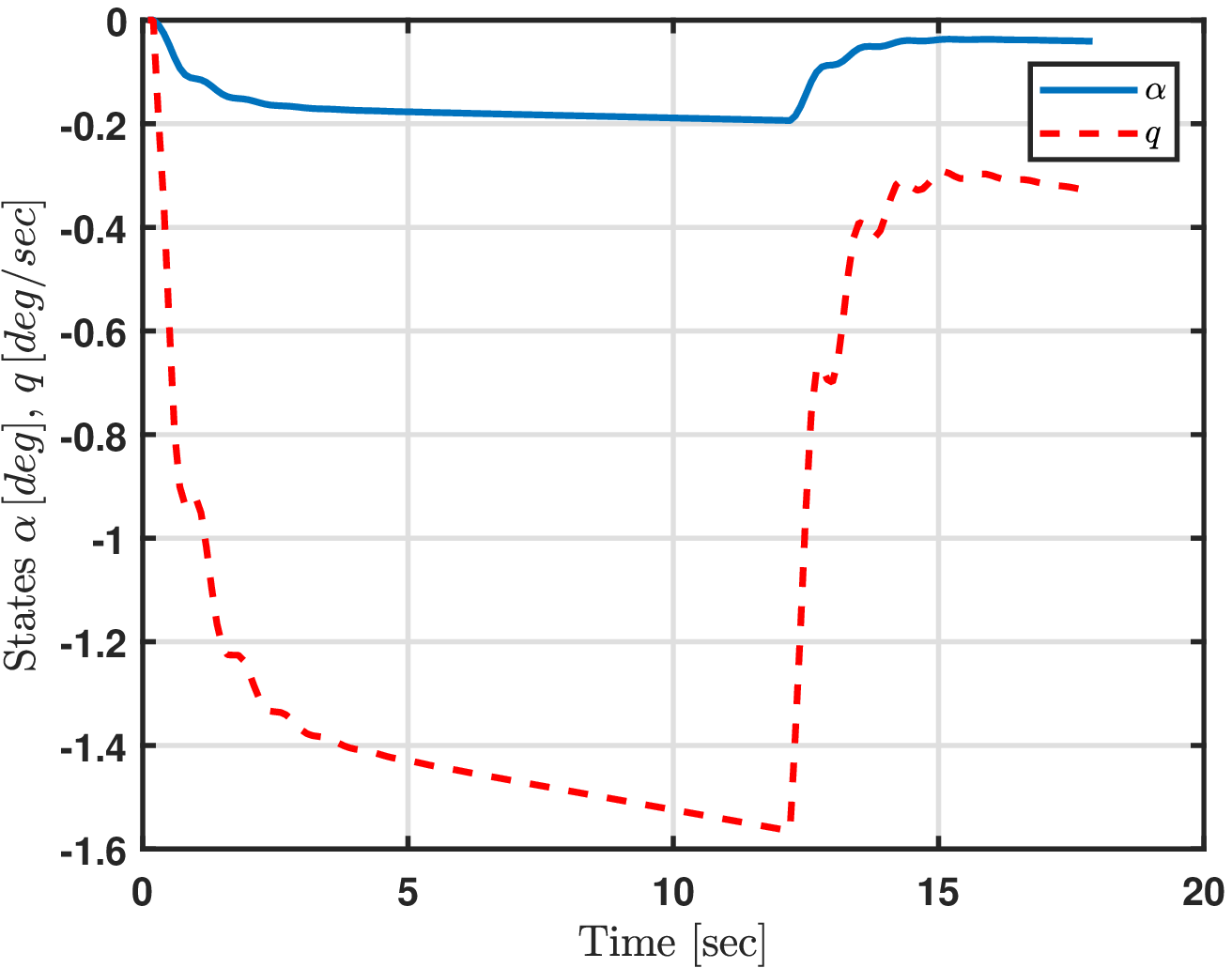}%
  }
  \\[2ex]
  \mbox{}\hfill
  \subcaptionbox{Elevator deflection~$\bm{u}(t)$%
    \label{fig:con_2}}
  {%
    \includegraphics[width=0.32\textwidth]{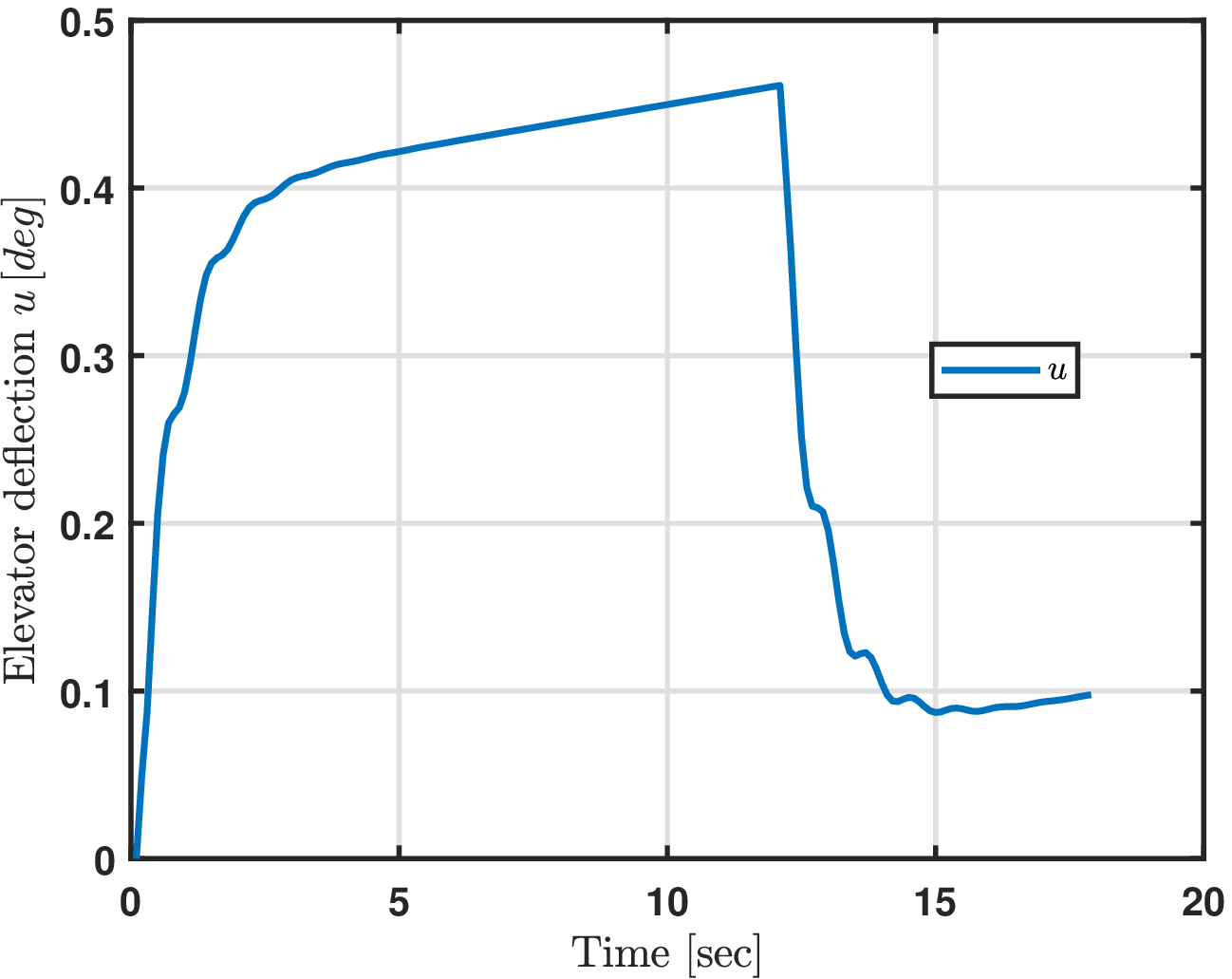}%
  }
  \hfill
  \subcaptionbox{Control gains~$K(t)$%
    \label{fig:act_2}}
  {%
    \includegraphics[width=0.32\textwidth]{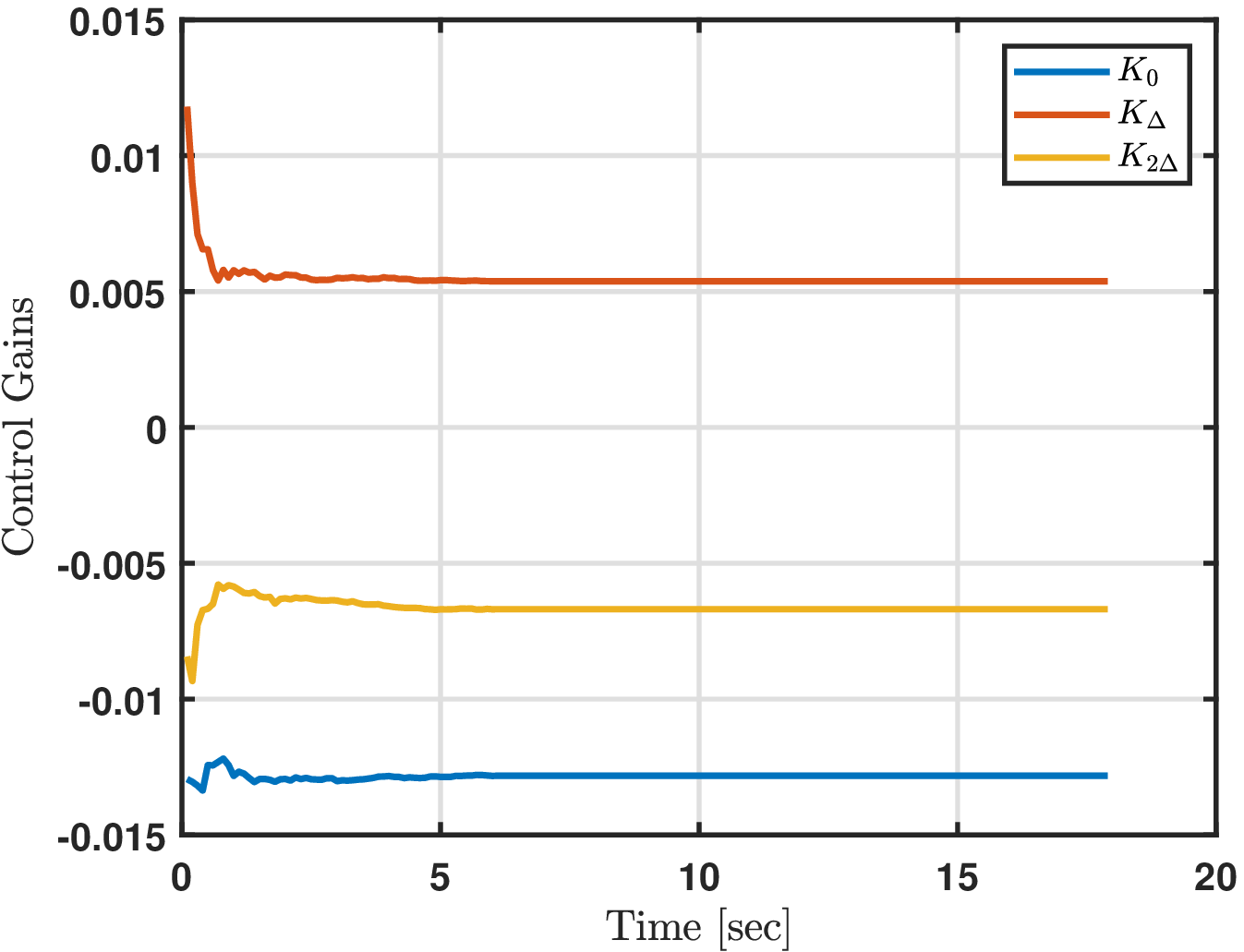}%
  }
  \hfill\mbox{}
  \caption{Simulation results of test case~2\label{fig:main 2}} 
\end{figure*}

The second test case challenges the performance of the adaptive learning controller by introducing a dynamic disturbance to the aircraft, such that $\dot{\bm{X}} = \bm{A} \, \bm{X}+ \bm{B} \, \left(\rho\,\bm{u}+\bm{\Xi}\,\bm{X}\right)$, where
\begin{align*}
  (\rho , \bm{\Xi})
  &
    =
    \begin{cases}
      \qty( 0.8052 , \vect{-0.2760 , -0.0858} ) \hspace*{-1em} & \text{, for } 1 \leq k < 60\\
      \qty( 0.5693 , \vect{-0.1959 , -0.0028} ) \hspace*{-1em} & \text{, for } 60 \leq k < 120\\
      \qty( 0.4187 , \vect{-0.5324 , -0.0002} ) \hspace*{-1em} & \text{, for } k \geq 120
    \end{cases}
\end{align*}
Furthermore, the controller is set to track a distorted reference signal compared to that used earlier. The results are demonstrated in \Cref{fig:main 2}. It is worth noticing how the error and the control gains converge in about 40 iterations (\SI{4}{\s}) in both test cases. 


\section{Conclusion}
\label{Conc}
The paper introduced an online adaptive learning control approach to solve a class of model-reference adaptive problems. The adaptive mechanism solves an integral Bellman equation using a reinforcement learning process. The solution does not rely on any information about the aircraft or the reference dynamics while using only the available measurements. The relation between the optimal control solution and the underlying adaptive control system is explained.  A Lyapunov approach is employed to show the stability features of the proposed adaptive learning control mechanism. The convergence properties of the reinforcement learning solution based on a Value Iteration process are highlighted and proven. The salient features of the proposed approach are demonstrated in the control of the longitudinal motion of an aircraft with unknown dynamics.

\IEEEtriggeratref{5} 
\IEEEtriggercmd{\enlargethispage{-2.7in}} 
\bibliographystyle{IEEEtran}
\bibliography{Bib/ref.bib}

\end{document}